\documentclass[aps,orb,twocolumn,amssymb]{revtex4}
\usepackage[dvipdf]{epsfig}
\usepackage{graphicx}
\begin{document}
\def\be{\begin{equation}}
\def\ee#1{\label{#1}\end{equation}}

\title{A model for a non-minimally coupled scalar field interacting with dark matter}

\author{J. B. Binder and G. M. Kremer\\
Departamento de F\'{\i}sica, Universidade Federal do Paran\'a\\
Caixa Postal 19044, 81531-990 Curitiba, Brazil}

\begin{abstract}

In this work we investigate the evolution of a Universe consisted of
a scalar field, a dark matter field and non-interacting baryonic
matter and radiation. The scalar field, which plays the role of dark
energy, is non-minimally coupled to space-time curvature, and drives
the Universe to a present accelerated expansion. The
non-relativistic dark matter field interacts directly with the dark
energy and has a pressure which follows from a thermodynamic theory.
We show that this model can reproduce the expected behavior of the
density parameters, deceleration parameter and luminosity distance.

\end{abstract}

 \maketitle
\section{Introduction}
Recently, well known astronomical observations with super-novae of
type Ia suggested that our Universe is presently submitted to an
accelerated expansion \cite{2,3} but the nature of the responsible
entity, the so-called dark energy, still remains unknown. The
simplest theoretical explanation for the acceleration consists in
introducing a cosmological constant and investigating the so-called
$\Lambda CDM$ model (see \cite{4}), which fits the present data very
well but has some important unsolved problems. Another possibility
is to introduce a minimally coupled scalar field $\phi \left(t
\right)$, which has been extensively studied by the scientific
community and could be invalidated if future observations imply that
the ratio between the pressure and the energy density of the scalar
field is restricted to values $p_{\phi} / \rho_{\phi} < -1$ (see
\cite{7}).

In this work we consider a scalar field non-minimally coupled to
space-time curvature, which was investigated in \cite{10} and widely
studied recently, among others, in the works
\cite{11,12,13,14,15,16,JK}. Also, aside several models for the dark
sector interaction \cite{17,dan,21}, we consider here a direct
coupling between dark matter and dark energy, following a model
proposed in \cite{18a}  and studied more recently in \cite{18} and
\cite{19}. We also use a thermodynamic theory~\cite{26} in order to
relate the effects of the non-minimally coupling to the dark matter
pressure~\cite{JK}. All components will be described by a set of
field equations, and the resulting observables -- namely, the
density parameters, the decelerating parameter and the luminosity
distance -- which are obtained as solutions of the field equations,
will be compared to the available data set in order to draw the
conclusions about the viability of this model. We show that
physically acceptable solutions are obtained and that there exist
some freedom parameters that will be important to fit the data from
incoming experiments. Units have been chosen so that $8 \pi G = c =
\hbar = k = 1$, whereas the metric tensor has signature ($+,-,-,-$).

\section{Field equations}

We shall  model the Universe as a mixture of a scalar field, a dark
matter field and non-interacting baryons and radiation. Furthermore,
we shall consider that the scalar field, which plays the role of
dark energy, is non-minimally coupled to the scalar
curvature~\cite{10} and that the dark matter interacts with the
scalar field according to a model proposed by
Wetterich~\cite{18a,18}.

 For an isotropic, homogeneous  and spatially flat Universe
described by the Robertson-Walker metric
 \be
  ds^2=dt^2-a(t)^2\left( dx^2+dy^2+dz^2 \right),
  \ee{1}
with $a(t)$ denoting the  cosmic scale factor, the Friedmann and
acceleration equations read
 \be
  H^2 = {\rho \over 3 \left( 1 - \xi \phi^2 \right)}, \qquad {\ddot a
\over a} = - {\rho+ 3p\over 6 \left( 1 - \xi \phi^2
 \right)},
 \ee{2}
 respectively. The quantities $\rho=\rho_b + \rho_r + \rho_{dm} +
\rho_\phi$ and $p=p_b + p_r + p_{dm} + p_\phi$ denote the energy
density and the pressure of the
 sources of the gravitational field, respectively, and they are given in terms of  the respective
 quantities for its constituents. Moreover, the dot denotes the differentiation with respect to time,
  $H=\dot a/a$ is the Hubble parameter,
 and $\xi$ the coupling constant between the scalar field $\phi$ and the curvature scalar.

The modified Klein-Gordon equation for the scalar field, which takes
into account the non-minimal coupling and the interaction with the
dark matter, reads
 \be
 \ddot \phi + 3 H \dot \phi + {d V \over d \phi}
+ {\xi \phi \over 1 - \xi \phi^2} \left( \rho - 3p \right) - \beta
\rho_{dm} = 0,
 \ee{3}
 where $\beta$ is a constant that couples the fields of dark energy and
 dark matter. The energy density and the pressure of the dark energy are given by
\begin{eqnarray} \label{4}
\rho_\phi &=& {1 \over 2}
 \dot \phi^2 + V + 6 \xi H \phi \dot \phi,  \\
\label{5} p_{\phi}&=& {1 \over 2} \dot \phi^2 - V - 2 \xi \left(
\phi \ddot  \phi + \dot \phi^2 + 2H \phi \dot \phi \right),
\end{eqnarray}
 respectively, where  $V( \phi)$ denotes the dark energy potential density. From
 (\ref{3}) -- (\ref{5}) it follows the evolution equation for the
 energy density of dark energy, namely
 \be
 \dot \rho_\phi + 3H \left( \rho_{\phi} + p_{\phi} \right) = - {2
 \xi \phi \dot \phi \rho \over 1 - \xi \phi^2} + \beta \rho_{dm} \dot
 \phi. \ee{6}
The first term on the right-hand side of (\ref{6}) is the
responsible for the energy transfer from the dark energy to the
gravitational field whereas the second refers to an energy  transfer
 to the dark matter field.

We shall assume that the baryons are non-relativistic particles so
that $p_b=0$ and that  the barotropic equation of state for the
radiation field $p_r=\rho_r/3$ holds. Hence, the evolution equations
for the energy densities of  non-interacting baryons and radiation
become
 \be
 \dot \rho_{b} + 3H\rho_{b}  =0,\qquad
 \dot \rho_{r} + 4H\rho_{r} =0,
 \ee{7}
respectively, and we get from (\ref{7}) the well-known
relationships: $\rho_b\propto 1/a^3$ and $\rho_r\propto 1/a^4$.

Now by using equations (\ref{2}) together with (\ref{6}) and
(\ref{7}) it follows the evolution equation for the energy density
of the dark matter field:
 \be \dot \rho_{dm} + 3H \left( \rho_{dm} + p_{dm}
\right) = -\beta \rho_{dm} \dot \phi.
 \ee{8}
Note that the term on the right-hand side of (\ref{8}) represents
the  energy transfer from the dark energy to the dark matter.

\section{Application}

Two different situations for the dark matter  could be analyzed,
namely, the one where it is considered as pressureless and the other
with a non-vanishing pressure. Here we shall restrict ourselves to
the latter case. In order to determine the expression for the dark
matter pressure, we recall that in the presence of matter creation
the first law of thermodynamics for adiabatic ($dQ=0$) open systems
reads~\cite{26}
 \be
 dQ=d(\rho V)+pdV-{\rho_{dm}+p_{dm}\over n_{dm}}d(n_{dm}V)=0,
 \ee{9}
where $V$ denotes the volume and it was supposed that only  dark
matter creation is  allowed. By considering $n_{dm} \propto
\rho_{dm}$ and $V\propto a^3$ it follows from (\ref{9}) together
with (\ref{2}) and (\ref{6}) -- (\ref{8}):
 \be
{2\xi\phi\dot\phi\rho\over 1-\xi\phi^2}=-{\rho_{dm}+p_{dm}\over
n_{dm}}(\dot n_{dm}+3Hn_{dm}).
 \ee{10}
 Now we use again
(\ref{8}) and obtain the following expression for the pressure of
the dark matter field
 \be
 {p_{dm} \over \rho_{dm}} = - {1 \over 2} -
{\beta \dot \phi \over 6H}  \pm \sqrt{\left( {1 \over 2} - {\beta
\dot \phi \over 6H} \right)^2\! \!+ {2 \xi \phi \dot \phi \rho \over
3 H \rho_{dm} \left( 1 - \xi \phi^2 \right)}}.
 \ee{11}
 We note  that only the plus sign must be chosen in the
above equation in order to obtain a positive dark matter pressure.
Moreover, we observe that for this case the dark matter pressure
vanishes when $\beta=\xi=0$.

We have solved numerically the system of differential equations
(\ref{3}) and (\ref{8}) as functions of the red-shift $z=1/a-1$ by
using (\ref{4}), (\ref{5}), (\ref{11}) together with $\rho_b\propto
1/a^3$ and $\rho_r\propto 1/a^4$ and the usual exponential potential
$V = V_0 e^{-\alpha \phi}$, where $\alpha$ and $V_0$ are constants.
Furthermore, the initial conditions for the energy densities were
chosen from the present known values given in the literature
\cite{25} for the density parameters $ \Omega_{i}(z) = {\rho_i(z) /
\rho(z)}$, i.e., $\Omega_{b}(0) = 0.04995$, $\Omega_{r}(0) =
5\times10^{-5}$ and $\Omega_{dm}(0) = 0.23$. The constant $V_0$ was
determined from the present value for the dark energy density
parameter, i.e., $\Omega_{\phi}(0)= 0.72$.

\begin{figure}\begin{center}
\includegraphics[width=6.5cm]{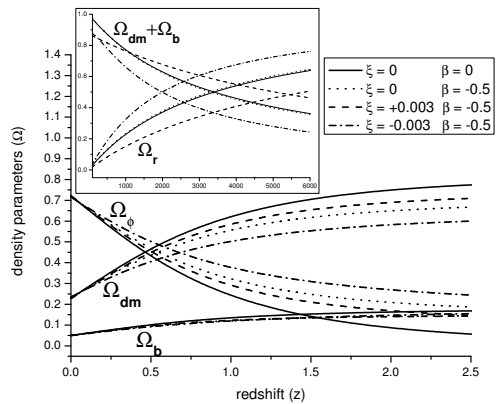}
\caption{Evolution of density parameters vs. red-shift.}
\end{center}\end{figure}

In Fig. 1 we have plotted the red-shift evolution of the density
parameters where it was considered a fixed value for the exponent
$\alpha$, namely $\alpha = 0.1$, and different values for the
coupling constants $\beta$ and $\xi$. From this figure we infer that
all kinds of coupling delay the growth of dark matter energy density
and the decay of dark energy density with respect to the red-shift,
by a comparison with the uncoupled case, i.e.,  $\beta=\xi=0$. Also,
positive values of $\xi$ imply that the radiation-matter (dark
matter plus baryons) equality takes place at higher red-shifts,
whereas negative values of $\xi$ at lower ones.

\begin{figure}\begin{center}
\includegraphics[width=6.5cm]{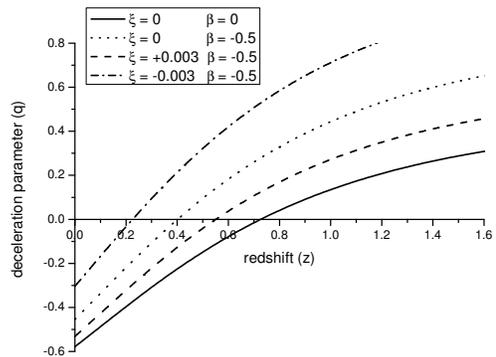}
\caption{Deceleration parameter vs. red-shift.}
\end{center}\end{figure}

The deceleration parameter $q = - {\ddot a a / \dot a ^2}$ as a
function of the red-shift is plotted in Fig. 2. According to current
experimental values \cite{6} the decelerated-accelerated transition
occurs at $z_T = 0.46 \pm 0.13$, whereas the present value  of the
deceleration parameter for  the $\Lambda CDM$ case  is
$q_0\approx-0.55$. We note from this figure that the
decelerated-accelerated transition shifts from $z_T \approx 0.7$ for
the uncoupled case ($\beta=\xi=0$) to $z_T \approx 0.2$  for $\beta
= -0.5$ and $\xi=0$. Moreover, by keeping a fixed value for $\beta$,
positive values for  $\xi$ cause later transitions  where negative
values for  $\xi$ imply earlier transitions. This means that, by
adjusting the coupling constants in this model, it is possible to
have concordance with the experimental data.

\begin{figure}\begin{center}
\includegraphics[width=6.5cm]{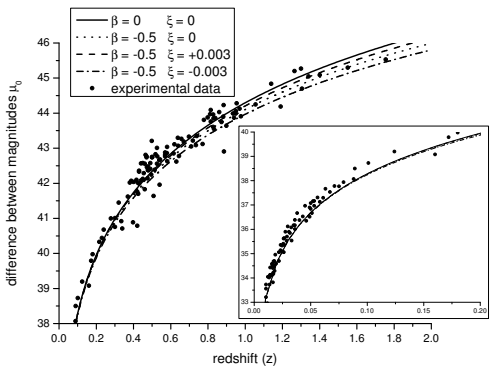}
\caption{Difference $\mu_0$ vs. red-shift.}
\end{center}\end{figure}

In Fig. 3 we have plotted the difference $\mu_0$ between the
apparent magnitude $m$ and the absolute magnitude $M$ of a source as
a function of the red-shift  where the circles represent the
experimental values taken from the work by Riess \emph{et al.}
\cite{6} for 185 data points of super-novae of type Ia. The
expression for $\mu_0$ reads $\mu_0 = m - M = 5 \log d_L + 25$,
where $d_L$ is the luminosity distance given in Mpc: $ d_L =( 1+z)
cH_0^{-1} \int^z_0 {dz / H (z)}.$  We infer from the figure that all
cases fit well  the experimental data at low red-shifts, but when
$\xi$ becomes more negative, the curve goes away from the  expected
experimental values.

\begin{figure}\begin{center}
\includegraphics[width=6.5cm]{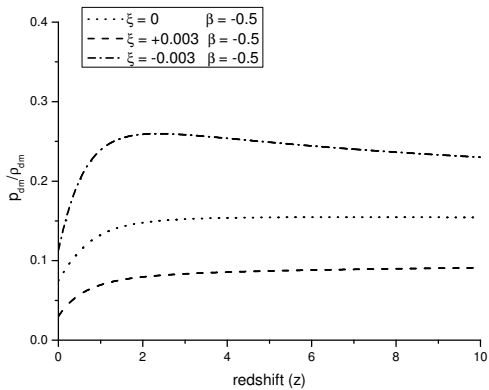}
\caption{Dark matter pressure vs. red-shift.}
\end{center}\end{figure}

The pressure of the dark matter as function of the red-shift is
plotted in Fig. 4. We infer from this figure that for a fixed value
of the coupling constant $\beta$, negative values for $\xi$ lead to
higher pressure of the dark matter field, and by increasing $\xi$ to
positive values the pressure decreases. In all cases the dark matter
pressure tend to a constant value for higher red-shifts.

\section{Conclusions}

The model analyzed here includes two kinds of couplings, namely, the
coupling of the scalar field with the curvature scalar and the
coupling of the dark matter field with the scalar field. Like  the
model investigated in the work~\cite{JK}, it has a strong dependence
on the initial conditions, since we have solved the  system of
coupled differential equations by imposing that the slope of the
scalar field has  small positive values at very high red-shifts,
which was a necessary condition to have positive values for the dark
energy density. However, as we have shown, there exist some freedom
in this model because the two coupling constants  can be adjusted in
order to fit present and future experimental data.


\end{document}